\documentclass[preprint,12pt]{elsarticle}
\usepackage{algorithm2e}
\usepackage[utf8]{inputenc}
\usepackage{graphicx,amssymb,amsmath,tikz,tcolorbox,amsthm,lineno,url,fullpage}
\usepackage[]{algorithm2e}

\usepackage{url}

\newenvironment{claimproof}{\paragraph{Proof of the claim:}}{\hfill$\triangleleft$}

\date{}

\newtheorem{corollary}{Corollary}

\newtheorem{lemma}{Lemma}
\newtheorem{theorem}{Theorem}

\newtheorem{ex}{Example}
\usepackage{thmtools}
\usepackage{thm-restate}

\usepackage{todonotes}

\usepackage{hyperref}
\usepackage{xcolor}
\hypersetup{
    colorlinks=true,
    linkcolor=blue,
    urlcolor=blue,
    citecolor=blue
}
\usepackage{cleveref}

\DeclareMathOperator{\rad}{rad}
\DeclareMathOperator{\MP}{mp}

\begin{document}
\begin{frontmatter}

\title{Relation between broadcast domination  and multipacking numbers on chordal and other hyperbolic graphs\tnoteref{label1}}
\tnotetext[label1]{A preliminary version of this paper was published in the proceedings of the CALDAM 2023 conference~\cite{DFIM23}.
}

\author[1]{Sandip Das}
\ead{sandipdas@isical.ac.in}

\author[2]{Florent Foucaud}
\ead{florent.foucaud@uca.fr}

\author[1]{Sk Samim Islam\corref{cor1}}
\ead{samimislam08@gmail.com}

\author[3]{Joydeep Mukherjee}
\ead{joydeep.m1981@gmail.com}

\address[1]{Indian Statistical Institute, Kolkata, India}
\address[2]{Université Clermont Auvergne, CNRS, Clermont Auvergne INP, Mines Saint-\'Etienne, LIMOS, 63000 Clermont-Ferrand, France}
\address[3]{Ramakrishna Mission Vivekananda Educational and Research Institute, Howrah, India}

\cortext[cor1]{Corresponding author}


\begin{abstract}
  For a graph $ G = (V, E) $ with a vertex set $ V $ and an edge set $ E $, a function
	$ f : V \rightarrow \{0, 1, 2, . . . , diam(G)\} $ is called a \emph{broadcast} on $ G $. For each
	vertex $ u \in V  $, if  there exists a vertex $ v $ in $ G $ (possibly, $ u = v $) such that $ f (v) > 0 $ and
	$ d(u, v) \leq f (v) $, then $ f $ is called a dominating broadcast on $ G $.  The cost of the dominating broadcast $f$ is the quantity $ \sum_{v\in V}f(v) $. The minimum cost of a
	dominating broadcast is the broadcast domination number of $G$, denoted by $ \gamma_{b}(G) $. 
	    
	    A multipacking is a set $ S \subseteq V  $ in a
	graph $ G = (V, E) $ such that for every vertex $ v \in V $ and for every integer $ r \geq 1 $, the
	ball of radius $ r $ around $ v $ contains at most $ r $ vertices of $ S $, that is, there are at most
	$ r $ vertices in $ S $ at a distance at most $ r $ from $ v $ in $ G $. The
	multipacking number of $ G $ is the maximum cardinality of a multipacking of $ G $ and
	is denoted by $ \MP(G) $.
	
	 It is known that $\MP(G)\leq \gamma_b(G)$ and that $\gamma_b(G)\leq 2\MP(G)+3$ for any graph $G$, and it was shown that $\gamma_b(G)-\MP(G)$ can be arbitrarily large for connected graphs. 
     For strongly chordal graphs, it is known that $\MP(G)=\gamma_b(G)$ always holds.
  
  We show that, for any connected chordal graph $G$, $\gamma_{b}(G)\leq \big\lceil{\frac{3}{2} \MP(G)\big\rceil}$. We also show that $\gamma_b(G)-\MP(G)$ can be arbitrarily large for connected chordal graphs by constructing an infinite family of connected chordal graphs such that the ratio $\gamma_b(G)/\MP(G)=10/9$, with $\MP(G)$ arbitrarily large. This result shows that, for chordal graphs, we cannot improve the bound $\gamma_{b}(G)\leq \big\lceil{\frac{3}{2} \MP(G)\big\rceil}$ to a bound in the form $\gamma_b(G)\leq c_1\cdot \MP(G)+c_2$, for any constant $c_1<10/9$ and $c_2$. Moreover, we show that $\gamma_{b}(G)\leq \big\lfloor{\frac{3}{2} \MP(G)+2\delta\big\rfloor} $ holds for all $\delta$-hyperbolic graphs. In addition, we provide a polynomial-time algorithm to construct a multipacking of a $\delta$-hyperbolic graph $G$ of size at least $  \big\lceil{\frac{2\MP(G)-4\delta}{3} \big\rceil} $.

\end{abstract}

\end{frontmatter}

\section{Introduction}
\label{sec:Introduction}

Covering and packing problems are fundamental in graph theory and algorithms~\cite{cornuejols2001combinatorial}. In this paper, we study two dual covering and packing problems called \emph{broadcast domination} and \emph{multipacking}. The broadcast domination problem has a natural motivation in telecommunication networks: imagine a network with radio emission towers, where each tower can broadcast information at any radius $r$  for a cost of $r$. The goal is to cover the whole network by minimizing the total cost. The multipacking problem is its natural packing counterpart and generalizes various other standard packing problems. Unlike many standard packing and covering problems, these two problems involve arbitrary distances in graphs, which makes them challenging. The goal of this paper is to study the relation between these two parameters in the class of chordal graphs and $\delta$-hyperbolic graphs.


Broadcast domination can be seen as a generalization of the classic dominating set problem. 
For a graph $ G = (V, E) $ with a vertex set $ V $, an edge set $ E $ and the diameter $diam(G)$, a function
	$ f : V \rightarrow \{0, 1, 2, . . . , diam(G)\} $ is called a \textit{broadcast} on $ G $. Suppose $G$ is a graph with a broadcast $f$. Let $d(u,v)$ be  the length of a shortest path joining the vertices $u$ and $v$ in $G$.  We say $v\in V$ is a \textit{tower} of $G$ if $f(v)>0$.	
Suppose $u, v  \in  V$ (possibly, $ u = v $) such that $ f (v) > 0 $ and
	$ d(u, v) \leq f (v) $, then we say $v$ \textit{broadcasts} (or  \textit{dominates}) $u$ and $u$ \textit{hears} a broadcast from $v$.

For each
	vertex $ u \in V  $, if  there exists a vertex $ v $ in $ G $ (possibly, $ u = v $) such that $ f (v) > 0 $ and
	$ d(u, v) \leq f (v) $, then $ f $ is called a \textit{dominating broadcast} on $ G $.
The \textit{cost} of the broadcast $f$ is the quantity $ \sigma(f)  $, which is the sum
	of the weights of the broadcasts over all vertices in $ G $. So, $\sigma(f)=  \sum_{v\in V}f(v)$. The minimum cost of a dominating broadcast in G (taken over all dominating broadcasts)  is the \textit{broadcast domination number} of G, denoted by $ \gamma_{b}(G) $.  So, $ \gamma_{b}(G) = \displaystyle\min_{f\in D(G)} \sigma(f)= \displaystyle\min_{f\in D(G)} \sum_{v\in V}f(v)$, where $D(G)=$ set of all dominating broadcasts on $G$. 
	
	Suppose $f$ is a dominating broadcast with $f(v)\in \{0,1\}$ $\forall v\in V(G)$, then $\{v\in V(G):f(v)=1\}$ is a \textit{dominating set} on $G$. The minimum cardinality of a dominating set is the \textit{domination number} which is denoted by $ \gamma(G) $.

An \textit{optimal broadcast} or \textit{optimal dominating broadcast} on a graph $G$ is a dominating broadcast with a cost equal to $ \gamma_{b}(G) $.	
A dominating broadcast is \textit{efficient} if no vertex
hears a broadcast from two different vertices. So, no tower can hear a broadcast from another tower in an efficient broadcast. There is a theorem that says,	for every graph there is an optimal dominating broadcast that is also efficient \cite{dunbar2006broadcasts}. 
	Define the \emph{ball} of radius $r$ around $v$ by $N_r[v]=\{u\in V(G):d(v,u)\leq r\}$.  Suppose $V(G)=\{v_1,v_2,v_3,\dots,v_n\}$. Let $c$ and $x$ be the vectors indexed by $(i,k)$ where $v_i\in V(G)$ and $1\leq k\leq diam(G)$, with the  entries $c_{i,k}=k$ and $x_{i,k}=1$ when $f(v_i)=k$ and $x_{i,k}=0$ when $f(v_i)\neq k$. Let $A=[a_{j,(i,k)}]$ be a matrix with the entries 
\begin{equation*}	a_{j,(i,k)}=
    \begin{cases}
        1 & \text{if }  v_j\in N_k[v_i]\\
        0 & \text{otherwise. }
    \end{cases} 
 \end{equation*}

	Hence, the broadcast domination number can be expressed as an integer linear program:  $$\gamma_b(G)=\min \{c.x :  Ax\geq \mathbf{1}, x_{i,k}\in \{0,1\}\}.$$	
	
    	A \textit{multipacking} is a set $ M \subseteq V  $ in a
	graph $ G = (V, E) $ such that   $|N_r[v]\cap M|\leq r$ for each vertex $ v \in V(G) $ and for every integer $ r \geq 1 $.	 The \textit{multipacking number} of $ G $ is the maximum cardinality of a multipacking of $ G $ and it
	is denoted by $ \MP(G) $. A \textit{maximum multipacking} is a multipacking $M$  of a graph $ G  $ such that	$|M|=\MP(G)$. If $M$ is a multipacking, we define   a vector $y$ with the entries $y_j=1$ when $v_j\in M$ and $y_j=0$ when $v_j\notin M$.  So, $$\MP(G)=\max \{y.\mathbf{1} :  yA\leq c, y_{j}\in \{0,1\}\}.$$  The \textit{maximum multipacking problem} is the dual integer program of the \textit{optimal dominating broadcast problem}.
	


\vspace{0.25cm}
\noindent\textbf{Brief Survey:}
 Erwin~\cite{erwin2004dominating,erwin2001cost} introduced broadcast domination in his doctoral thesis in
2001. Multipacking was  introduced in Teshima’s Master’s Thesis \cite{teshima2012broadcasts} in 2012 (also see \cite{beaudou2019multipacking,cornuejols2001combinatorial,dunbar2006broadcasts,meir1975relations}). 
For general graphs, surprisingly, an optimal dominating broadcast can be found in polynomial-time $O(n^6)$~\cite{heggernes2006optimal}. The same problem can be solved in linear time for trees~\cite{brewster2019broadcast}. However, until now, there is no known polynomial-time algorithm to find a maximum multipacking of general graphs (the problem is also not known to be NP-hard). However, polynomial-time algorithms are known for trees and more generally, strongly chordal graphs~\cite{brewster2019broadcast}. See~\cite{foucaud2021complexity} for other references concerning algorithmic results on the two problems.

We know that $\MP(G)\leq \gamma_b(G)$, since broadcast domination and multipacking are dual problems~\cite{brewster2013new}, and $\gamma_b(G)\leq \rad(G)$~\cite{erwin2001cost}, where $\rad(G)$ is the radius of $G$. Moreover, it is known that $\gamma_b(G)\leq 2\MP(G)+3$~\cite{beaudou2019broadcast} and it is a conjecture that $\gamma_b(G)\leq 2\MP(G)$ for every graph $G$~\cite{beaudou2019broadcast}.  
If a graph $G$ has diameter~2 but no universal vertex, then $\gamma_b(G)=2$ (since no single vertex can dominate all other vertices using a broadcast of cost~1, but every vertex can reach every other vertex by broadcasting at distance~2) and $\MP(G)=1$ (since any two vertices are either adjacent or have a common neighbour). Examples of such graphs are the cycles $C_4$, $C_5$, complete bipartite graphs that are not stars, the Petersen graph, the Wagner graph, the Grötzsch graph, the Clebsch graph, etc. One can also take such a graph and blow up some of its vertices into independent sets or cliques to obtain more graphs. Thus, the conjectured upper bound of $2$ for the ratio $\frac{\gamma_b(G)}{\MP(G)}$, would be optimal for infinitely many graphs. However, no family of connected graphs with arbitrarily large value of their multipacking numbers is known that reach this ratio (in fact, no such graph is known with multipacking number~3 or more~\cite{beaudou2019broadcast,hartnell2014difference}).


Hartnell and Mynhardt~\cite{hartnell2014difference} constructed a family of connected graphs such that the difference $\gamma_b(G)-\MP(G)$ can be arbitrarily large and in fact, for which the ratio $\gamma_b(G)/ \MP(G)=4/3$. Until recently, this was the best known construction. After the submission of the current paper, Rajendraprasad, Sani, Sasidharan and Sen~\cite{Rajendraprasad25} showed that hypercubes form an interesting example family in this regard. Indeed, it was already known that for the hypercube $H_d$ of dimension $d$, we have $\gamma_b(H_d)=d-1$~\cite{bresar2009broadcast}, and the above authors~\cite{Rajendraprasad25} showed that $\MP(H_d)\leq \frac{d}{2}+6\sqrt{2d}$. Therefore, for general connected graphs, 
$$\lim_{\MP(G)\to \infty}\sup\Bigg\{\frac{\gamma_{b}(G)}{\MP(G)}\Bigg\}= 2.$$ 
A natural question comes to mind: What is the maximum value of this ratio for other classes of connected graphs? It is known that $\gamma_b(G)=\MP(G)$ holds whenever $G$ is a strongly chordal graph~\cite{brewster2019broadcast}. Thus, a natural class to study is the class of chordal graphs. Note that the construction by Hartnell and Mynhardt~\cite{hartnell2014difference} as well as hypercubes, which provide known examples with a large ratio, are not chordal.

\vspace{0.25cm}
\noindent\textbf{Our Contribution:}
A \textit{chordal graph} is an undirected simple graph in which all cycles of four or more vertices have a chord, which is an edge that is not part of the cycle but connects two vertices of the cycle.  Chordal graph is a superclass of interval graph. In this paper, we study the multipacking problem of chordal graphs.

\begin{figure}[t]
\centering
\scalebox{0.9}{\begin{tikzpicture}[node distance=7mm]

\tikzstyle{mybox}=[fill=white,line width=0.5mm,rectangle, minimum height=.8cm,fill=white!70,rounded corners=1mm,draw];
\tikzstyle{myedge}=[line width=0.5mm]
\newcommand{\tworows}[2]{\begin{tabular}{c}{#1}\\{#2}\end{tabular}}

 \node[mybox] (hyper)  {bounded hyperbolicity};
    \node[mybox] (treelength)  [below=of hyper,yshift=3mm]  {bounded treelength} edge[myedge] (hyper);
    \node[mybox] (chordality) [below=of treelength,yshift=3mm] {bounded chordality} edge[myedge] (treelength);
      \node[mybox] (diam) [below left=of treelength,xshift=-10mm,yshift=1mm] {bounded diameter} edge[myedge] (treelength);
    \node[mybox] (ATfree) [below right=of chordality,xshift=10mm,yshift=1mm] {asteroidal triple-free} edge[myedge, fill=red!30] (chordality);
    \node[mybox] (cocomp) [below =of ATfree,yshift=3mm] {co-comparability} edge[myedge] (ATfree);
    \node[mybox] (chordal) [below =of chordality,yshift=3mm] {chordal} edge[myedge] (chordality);
    \node[mybox, fill=gray!30] (strongly) [below =of chordal,yshift=3mm] {strongly chordal} edge[myedge] (chordal);
    \node[mybox] (split) [left =of strongly,xshift=-3mm] {split} edge[myedge] (chordal) edge[myedge] (diam);
    \node[mybox, fill=gray!30] (interval) [below right=of strongly,yshift=1mm] {interval} edge[myedge] (strongly) edge[myedge] (cocomp);
    \node[mybox] (perm) [below right=of cocomp,yshift=1mm] {permutation} edge[myedge] (cocomp);
    \node[mybox, fill=gray!30] (block) [below =of strongly,yshift=1mm]
    {block} edge[myedge] (strongly);
    \node[mybox, fill=gray!30] (tree) [below=of block,yshift=3mm] {trees} edge[myedge] (block);
  \end{tikzpicture}}

\caption{Inclusion diagram for graph classes mentioned in this paper (and related ones). If a class $A$ has an upward path to class $B$, then $A$ is included in $B$. For the graphs in the gray classes, the broadcast domination number is equal to the multipacking number, but this is not true for the white classes. 
}
\label{fig:diagram}
\end{figure}
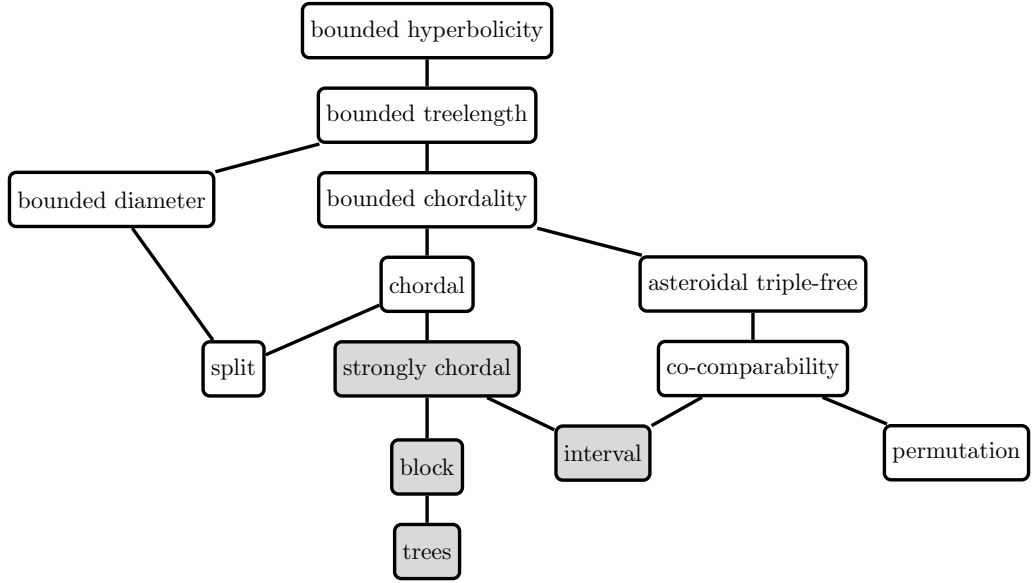

We also generalize this study as follows. Gromov~\cite{gromov1987hyperbolic} introduced $\delta$-hyperbolicity measures in order to study geometric group theory via Cayley
graphs. 
Let $d$ be the shortest-path metric of a graph $G$. The graph $G$ is called a \emph{$\delta$-hyperbolic graph} if for any four
vertices $u, v, w, x \in V(G)$, the two larger of the three sums $d(u, v) + d(w, x)$, $d(u, w) + d(v, x)$, $d(u, x) +
d(v, w)$ differ by at most $2\delta$. A graph class $\mathcal{G}$ is said to be hyperbolic if there exists a constant $\delta$ such that every graph $G \in \mathcal{G}$ is $\delta$-hyperbolic. Trees are $0$-hyperbolic graphs~\cite{buneman1974note}, and chordal graphs are 1-hyperbolic~\cite{brinkmann2001hyperbolicity}. In general, hyperbolicity is a measurement of the deviation of the distance function of a graph from a tree metric. Many other interesting graph
classes including 
co-comparability graphs \cite{corneil2013ldfs}, asteroidal-triple free graphs \cite{corneil1997asteroidal}, permutation graphs \cite{golumbic2004algorithmic}, 
graphs with bounded chordality or treelength are hyperbolic \cite{chepoi2008diameters,keil2017algorithm}. See Figure~\ref{fig:diagram} for a diagram representing the inclusion hierarchy between these classes. Moreover, hyperbolicity is a measure that captures properties of real-world graphs such as
the Internet graph \cite{shavitt2004curvature} or database relation graphs \cite{walter2002interactive}. 
In this paper, we study the multipacking problem of hyperbolic graphs.

We start by bounding the multipacking number of a chordal graph: 

\begin{restatable}{proposition}{multipackingbroadcastrelation}
\label{prop:gammabGleq3/2mpG}
 If $G$ is a connected chordal graph, then $\gamma_{b}(G)\leq \big\lceil{\frac{3}{2} \MP(G)\big\rceil} $. 
\end{restatable} 

The question of the algorithmic complexity of computing the multipacking number of a graph  
has been repeatedly addressed by numerous authors, yet it has persisted as an unsolved challenge for the past decade. However, polynomial-time algorithms are known for trees and more generally, strongly chordal graphs~\cite{brewster2019broadcast}. Even for trees, the algorithm for finding a maximum multipacking is very non-trivial. Chordal graphs form a superclass of the class of strongly chordal graphs. In this paper, we provide a $(\frac{3}{2}+o(1))$-approximation algorithm to find a multipacking on chordal graphs.   

\begin{restatable}{proposition}{approximationalgorithm}\label{prop:3/2mpGapprox} If $G$ is a connected chordal graph, there is a polynomial-time algorithm to construct a multipacking of $G$ of size at least $  \big\lceil{\frac{2\MP(G)-1}{3} \big\rceil}$.
\end{restatable}

This approximation algorithm is based on finding a diametral path which is almost double the radius for chordal graphs. The set of every third vertex on this path yields a multipacking.

Hartnell and Mynhardt~\cite{hartnell2014difference} constructed a family of connected graphs $H_k$ such that $\MP(H_k)=3k$ and $\gamma_b(H_k)=4k$. There is no such construction for trees and more generally, strongly chordal graphs, since multipacking number and broadcast domination number are the same for these graph classes~\cite{brewster2019broadcast}. In this paper, we construct a family of connected chordal graphs $H_k$ such that $\MP(H_k)=9k$ and $\gamma_b(H_k)=10k$ (Figure~\ref{fig:Names}).

\begin{restatable}{theorem}{multipackingbroadcastgape}\label{thm:9k10k}
 For each positive integer $k$, there is a connected chordal graph $H_k$ such that $\MP(H_k)=9k$ and $\gamma_b(H_k)=10k$.
\end{restatable}

Theorem \ref{thm:9k10k} directly establishes the following corollary.

\begin{restatable}{corollary}{gammabGdiffmpG}\label{cor:gammabG-mpG} The difference $ \gamma_{b}(G) -  \MP(G) $ can be arbitrarily large for connected chordal graphs.
\end{restatable}

We mentioned earlier that, for general connected graphs, the range of the value of the expression $\lim_{\MP(G)\to \infty}\sup\{\gamma_{b}(G)/\MP(G)\}$ is the interval $[4/3,2]$. We found the range of this expression for chordal graphs. Proposition \ref{prop:gammabGleq3/2mpG} and Theorem \ref{thm:9k10k} yield the following corollary.

\begin{restatable}{corollary}{gammabGbympG} \label{cor:gammabG/mpG} For connected chordal graphs $G$, 
$$\displaystyle\frac{10}{9}\leq\lim_{\MP(G)\to \infty}\sup\Bigg\{\frac{\gamma_{b}(G)}{\MP(G)}\Bigg\}\leq \frac{3}{2}.$$
\end{restatable}

We also make a connection with the \emph{fractional} versions of the two concepts dominating broadcast and multipacking, as introduced in~\cite{brewster2013broadcast}.

We establish a relation between broadcast domination and multipacking numbers of $\delta$-hyperbolic graphs using the same method  that
works for chordal graphs.  We state that in the following  proposition:

\begin{restatable}{proposition}{deltaMultipackingBroadcastRelation}\label{prop:delta_multipacking_broadcast_relation}
     If $G$ is a $\delta$-hyperbolic graph, then $\gamma_{b}(G)\leq \big\lfloor{\frac{3}{2} \MP(G)+2\delta\big\rfloor} $.

\end{restatable}

We used the same method that
works for chordal graphs to  provide an approximation algorithm to find a large multipacking of $\delta$-hyperbolic graphs.

\begin{restatable}{proposition}{approxdeltaMultipacking}\label{prop:approx_delta_multipacking}
    If $G$ is a $\delta$-hyperbolic graph, there is a polynomial-time algorithm to construct a multipacking of $G$ of size at least $  \big\lceil{\frac{2\MP(G)-4\delta}{3} \big\rceil} $.
\end{restatable}

A graph is \emph{$k$-chordal} (or, bounded chordality graph with bound $k$) if it does not contain any induced $n$-cycle for $n > k$. The class of $3$-chordal graphs is the class of usual chordal graphs. Brinkmann, Koolen and Moulton \cite{brinkmann2001hyperbolicity} proved that every chordal graph is $1$-hyperbolic. In 2011, Wu and Zhang \cite{wu2011hyperbolicity} established that a $k$-chordal graph is always a $\frac{\lfloor\frac{k}{2}\rfloor}{2}$-hyperbolic graph, for every $k\geq 4$. A graph $G$ has \emph{treelength} at most $d$ if it admits a tree-decomposition where the maximum distance between any two vertices in the same bag have distance at most $d$ in $G$~\cite{DG07}. Clearly, a graph of diameter $d$ has treelength at most $d$. Graphs of treelength at most $d$ have hyperbolicity at most $d$~\cite{chepoi2008diameters}. Moreover, $k$-chordal graphs have treelength at most $2k$~\cite{DG07}. 
Additionally, it is known that the hyperbolicity is $1$ for the graph classes: asteroidal-triple free, co-comparability and permutation graphs~\cite{wu2011hyperbolicity}. 
These results about hyperbolicity and Proposition \ref{prop:delta_multipacking_broadcast_relation} directly establish the following.

\begin{corollary}\label{cor:k-chordal} Let $G$ be a graph.

\noindent(i) If $G$ is an asteroidal-triple free, co-comparability or permutation graph, then $\gamma_{b}(G)\leq \big\lfloor{\frac{3}{2} \MP(G)+2}\big\rfloor $.

\noindent(ii) If $G$ is a $k$-chordal graph where $k\geq 4$, then $\gamma_{b}(G)\leq \big\lfloor{\frac{3}{2} \MP(G)+\lfloor\frac{k}{2}\rfloor}\big\rfloor $.

\noindent(iii) If $G$ has treelength $d$, then $\gamma_{b}(G)\leq \big\lfloor{\frac{3}{2} \MP(G)+2d}\big\rfloor $.
\end{corollary}

\vspace{0.25cm}
\noindent\textbf{Organisation:} In Section \ref{sec:Definitions and notation}, we recall some definitions and notations. 
In Section \ref{sec:An inequality linking Broadcast domination and Multipacking numbers}, we  establish a relation between multipacking and dominating broadcast on the chordal graphs. In the same section, we provide a $(\frac{3}{2}+o(1))$-factor approximation algorithm for finding multipacking on the same graph class. In Section \ref{sec:Unboundedness of the gap between Broadcast domination and Multipacking numbers of Chordal graphs}, we prove our main result which says that the difference $ \gamma_{b}(G) -  \MP(G) $ can be arbitrarily large for connected chordal graphs. In Section \ref{sec:A study of Broadcast domination and Multipacking numbers on Hyperbolic graphs},  we relate the broadcast domination and multipacking number of $\delta$-hyperbolic graphs and provide an approximation algorithm for the multipacking problem of the same. We conclude in Section \ref{sec:Conclusion}. 

\section{Definitions and notations}\label{sec:Definitions and notation}

 Let $G=(V,E)$ be a graph and $d_G(u,v)$ be the length of a shortest path joining two vertices $u$ and $v$ in  $G$, we simply write $d(u,v)$ when there is no confusion. Let $diam(G)=\max\{d(u,v):u,v\in V(G)\}$. Diameter (or diametral path) is a path of $G$  of the length  $diam(G)$. 
 $N_r[u]=\{v\in V:d(u,v)\leq r\}$ where $u\in V$. The \textit{eccentricity} $e(w)$  of a vertex $w$ is $\min \{r:N_r[w]=V\}$. The \textit{radius} of the graph $G$ is $\min\{e(w):w\in V\}$, denoted by $\rad(G)$.  The \textit{center set} $C(G)$ of the graph $G$  is the set of all vertices of minimum eccentricity, i.e., $C(G)=\{v\in V:e(v)=\rad(G)\}$. Each vertex in the set $C(G)$ is called a \textit{center} of the graph $G$.

\section{An inequality linking Broadcast domination and Multipacking numbers of Chordal Graphs}\label{sec:An inequality linking Broadcast domination and Multipacking numbers}

In this section, we use results from the literature to show that the general bound connecting multipacking number and broadcast domination number can be improved for chordal graphs.

\begin{theorem}[\cite{hartnell2014difference}]\label{d+1/3leqmpG}
If $G$ is a connected graph of order at least 2  having  diameter $ d $ and  multipacking number  $ \MP(G) $, where $P=v_0,\dots,v_d$ is a diametral  path of $G$, then the set $M=\{v_i:i\equiv 0 \text{ } (mod \text{ } 3), i=0,1,\dots,d\}$ is a multipacking of $G$ of size $\big\lceil{\frac{d+1}{3}\big\rceil}$ and  $\big\lceil{\frac{d+1}{3}\big\rceil}\leq \MP(G)$.
\end{theorem}

\begin{theorem}[\cite{erwin2001cost,teshima2012broadcasts}]  \label{mpGleqgammabG}  If $G$ is a connected graph of order at least 2  having radius $ r $, diameter $ d $, multipacking number  $\MP(G) $, broadcast domination number $ \gamma_{b}(G) $ and domination number $\gamma(G)$, then $ \MP(G)\leq \gamma_{b}(G) \leq min\{\gamma (G),r\}$.
\end{theorem} 

\begin{theorem}[\cite{laskar1983powers}] \label{2rleqd+2} If $ G $ is a connected chordal graph with radius $ r $ and diameter $ d $, then $ 2r\leq d+2 $.
\end{theorem} 

Using these results we prove the following proposition:

\multipackingbroadcastrelation*

\begin{proof}
From Theorem \ref{d+1/3leqmpG}, $\big\lceil{\frac{d+1}{3}\big\rceil}\leq \MP(G)$  which implies that $d \leq 3\MP(G)-1$. Moreover, from  Theorem \ref{mpGleqgammabG} and Theorem \ref{2rleqd+2},  $\gamma_{b}(G)\leq r \leq \big\lfloor{\frac{d+2}{2}\big\rfloor} \leq \big\lfloor{\frac{(3\MP(G)-1)+2}{2}\big\rfloor}$ $=\big\lfloor{\frac{3}{2} \MP(G)+\frac{1}{2}\big\rfloor}  $. Therefore, $\gamma_{b}(G)\leq \big\lfloor{\frac{3}{2} \MP(G)+\frac{1}{2}\big\rfloor} = \big\lceil{\frac{3}{2} \MP(G)\big\rceil}$. 
\end{proof}

The proof of Proposition~\ref{prop:gammabGleq3/2mpG} has the following algorithmic application.

\approximationalgorithm*

\begin{proof} If $P=v_0,\dots,v_d$ is a diametrical path of $G$, then the set $M=\{v_i:i\equiv 0 \text{ } (mod \text{ } 3), i=0,1,\dots,d\}$ is a multipacking of $G$ of size $\big\lceil{\frac{d+1}{3}\big\rceil}$ by Theorem \ref{d+1/3leqmpG}. We can construct $M$ in polynomial-time since we can find a diametral path of a graph $G$ in polynomial-time. Moreover, from  Theorem \ref{d+1/3leqmpG}, Theorem \ref{mpGleqgammabG} and Theorem \ref{2rleqd+2},  $\big\lceil{\frac{2\MP(G)-1}{3}\big\rceil}\leq\big\lceil{\frac{2r-1}{3}\big\rceil}\leq\big\lceil{\frac{d+1}{3}\big\rceil}\leq \MP(G)$. 
\end{proof}


\begin{figure}[h]
    \centering
   \includegraphics[height=3.1cm]{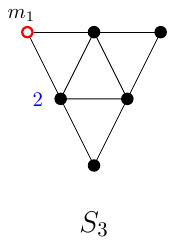}
    \caption{$S_3$ is a connected chordal graph   with $\gamma_b(S_3)=2$ and $\MP(S_3)=1$}
    \label{fig:S3}
\end{figure}

\begin{figure}[h]
    \centering
   \includegraphics[height=3.3cm]{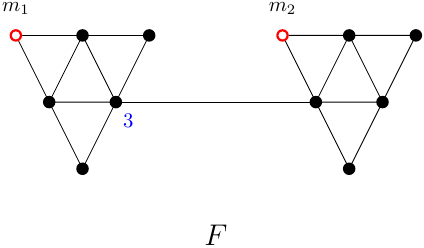}
    \caption{$F$  is a connected chordal graph   with $\gamma_b(F)=3$ and $\MP(F)=2$}
    \label{fig:F}
\end{figure}

\begin{figure}[h]
    \centering
   \includegraphics[height=3.2cm]{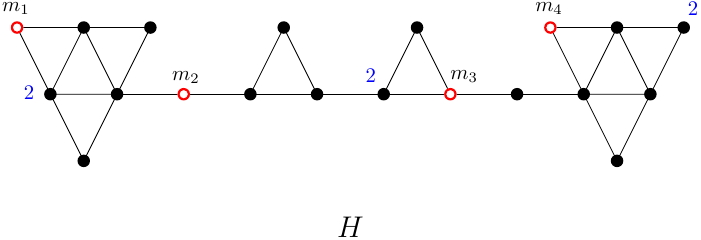}
    \caption{$H$ is a connected chordal graph  with $\gamma_b(H)=6$ and $\MP(H)=4$}
    \label{fig:H}
\end{figure}

Here we have some examples of graphs that achieve the equality of the bound in Proposition \ref{prop:gammabGleq3/2mpG}.

\begin{ex} The connected  chordal graph  $S_{3}$ (Figure~\ref{fig:S3}) has $  \MP(S_{3}) =1 $ and $\gamma_{b}(S_{3}) =2$. So, here  $ \gamma_{b}(S_{3}) = \big\lceil{\frac{3}{2} \MP(S_3)\big\rceil} $.
\end{ex}

\begin{ex} The  connected   chordal graph  $F$ (Figure~\ref{fig:F})  has $  \MP(F) =2 $ and $\gamma_{b}(F) =3$. So, here  $ \gamma_{b}(F) = \big\lceil{\frac{3}{2} \MP(F)\big\rceil} $.
\end{ex}

\begin{ex}
The  connected  chordal graph   $H$ (Figure~\ref{fig:H})  has $  \MP(H) =4 $ and $\gamma_{b}(H) =6$. So, here  $ \gamma_{b}(H) = \big\lceil{\frac{3}{2} \MP(H)\big\rceil} $.
\end{ex}

We could not find an example of  connected  chordal graph with $\MP(G)=3$ and $\gamma_{b}(G) =\big\lceil{\frac{3}{2} \MP(G)\big\rceil}=5$. 

\section{Unboundedness of the gap between Broadcast domination and Multipacking numbers of Chordal graphs}\label{sec:Unboundedness of the gap between Broadcast domination and Multipacking numbers of Chordal graphs}

In this section, our goal is to show that  the difference between broadcast domination number and multipacking number of connected chordal graphs  can be arbitrarily large. We prove this using the following theorem that we prove later.

\multipackingbroadcastgape*

Theorem \ref{thm:9k10k} yields the following.

\gammabGdiffmpG*

Moreover, Proposition \ref{prop:gammabGleq3/2mpG} and Theorem \ref{thm:9k10k} yield the following.

\gammabGbympG*

\subsection{Proof of Theorem \ref{thm:9k10k}}

\begin{figure}[h]
    \centering
   \includegraphics[height=3.3cm]{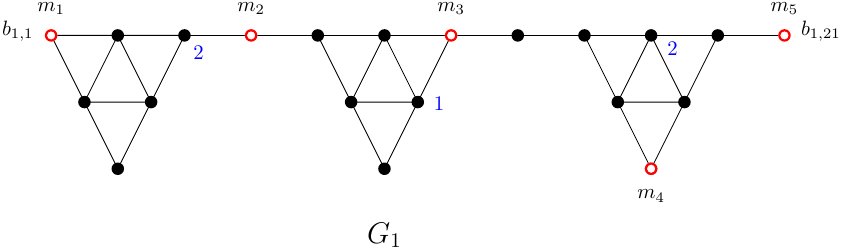}
    \caption{$G_1$ is a connected chordal graph  with $\gamma_b(G_1)=5$ and $\MP(G_1)=5$. $M_1=\{m_i:1\leq i \leq  5\}$ is a multipacking of size $5$.}
    \label{fig:G1m}
\end{figure}

\begin{figure}[h]
    \centering
   \includegraphics[height=7.1cm]{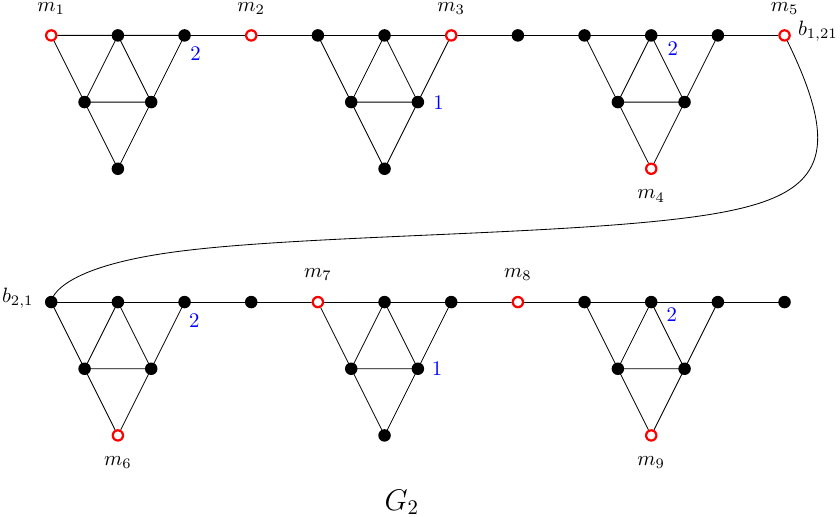}
    \caption{Graph $G_2$ with $\gamma_b(G_2)=10$ and $\MP(G_2)=9$. $M=\{m_i:1\leq i \leq  9\}$ is a multipacking of size $9$.}
    \label{fig:G2m}
\end{figure}

\begin{figure}[ht]
    \centering
   \includegraphics[height=11.3cm]{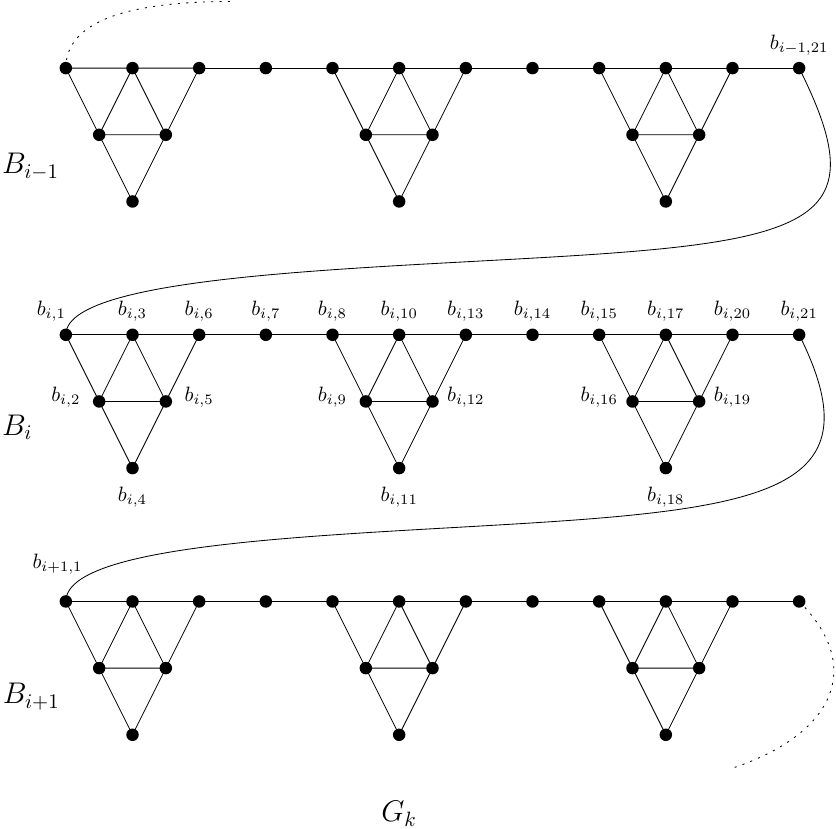}
    \caption{Graph $G_k$}
    \label{fig:Names}
\end{figure}

Consider the graph $G_{1}$ as in Figure \ref{fig:G1m}.  
Let $B_1$ and $B_2$ be two isomorphic copies of $G_1$. The vertices of $B_1$ and $B_2$ are labeled as the vertices of $B_i$ and $B_{i+1}$ (for $i=1$) in Fig~\ref{fig:Names}.  Join $b_{1,21}$ of  $B_1$ and $b_{2,1}$ of $B_2$ by an edge (Figure~\ref{fig:G2m} and \ref{fig:Names}).  We denote this new graph by $G_2$ (Figure~\ref{fig:G2m}). In this way, we form $G_k$ by joining $k$ isomorphic copies of $G_1$ : $B_1,B_2,\cdots,B_k$ (Figure~\ref{fig:Names}). Here $B_i$ is joined with $B_{i+1}$ by joining $b_{i,21}$ and $b_{i+1,1}$.  We say that $B_i$ is the $i$-th block of $G_k$.  $B_i$ is an induced subgraph of $G_k$ as given by $B_i= G_k[\{b_{i,j}:1\leq j \leq 21\}]$. Similarly, for $1\leq i\leq2k-1$, we define $B_i\cup B_{i+1}$, induced subgraph of $G_{2k}$, as $B_i\cup B_{i+1}=G_{2k}[\{b_{i,j},b_{i+1,j}:1\leq j \leq 21\}]$.  We prove Theorem \ref{thm:9k10k}  by establishing  that $\gamma_b(G_{2k})=10k$ and $\MP(G_{2k})=9k$. 

Our proof of Theorem \ref{thm:9k10k} is accomplished through a set of lemmas which are stated and proved below. We begin by observing a basic fact about multipacking in a graph. We formally state it in Lemma \ref{duv>=3} for ease of future reference.

\begin{lemma} \label{duv>=3} Suppose $M$ is a multipacking in a graph $G$. If $u,v \in M$ and $u \neq v$, then $d(u,v)\geq 3$.
\end{lemma}
\begin{proof}  If $d(u,v)=1$, then $u,v\in N_1[v]\cap M$, then $M$ cannot be a multipacking. So, $d(u,v)\neq 1$. If $d(u,v)=2$, then there exists a common neighbour $w$ of $u$ and $v$. So, $u,v\in N_1[w]\cap M$, then $M$ cannot be a multipacking. So, $d(u,v)\neq 2$. Therefore,  $d(u,v)>2$. 
\end{proof}

\begin{lemma}\label{mpG2k>=9k} $\MP(G_{2k})\geq 9k$, for each positive integer $k$.
\end{lemma}

\begin{proof}
Consider the set 
$M_{2k}=\{b_{2i-1,1},b_{2i-1,7},b_{2i-1,13},b_{2i-1,18},b_{2i-1,21},b_{2i,4}, $ $ b_{2i,8},b_{2i,14}, b_{2i,18} :1\leq i \leq k\}$ (Figure~\ref{fig:Names}) of size $9k$. 
 We want to show that $M_{2k}$ is a multipacking of $G_{2k}$. So, we have to prove that, $|N_r[v]\cap M_{2k}|\leq r$ for each vertex $ v \in V(G_{2k}) $ and for every integer $ r \geq 1 $. We prove this statement using induction on $r$.	It can be checked that $|N_r[v]\cap M_{2k}|\leq r$ for each vertex $ v \in V(G_{2k}) $ and for each  $ r \in \{1,2,3,4\} $. Now assume that the statement is true for $r=s$, we want to prove that, it is true for $r=s+4$. Observe that, $|(N_{s+4}[v]\setminus N_{s}[v])\cap M_{2k}|\leq 4$ for every vertex $ v\in V(G_{2k})$. Therefore,  $|N_{s+4}[v]\cap M_{2k}|\leq |N_{s}[v]\cap M_{2k}|+4\leq s+4$. So, the statement is true. Therefore, $M_{2k}$ is a multipacking of $G_{2k}$. So, $\MP(G_{2k})\geq |M_{2k}|= 9k$. 
\end{proof}

\begin{lemma} \label{mpG1=5}  $\MP(G_{1})=5$.
\end{lemma}
\begin{proof}  $V(G_1)=N_3[b_{1,7}]\cup N_2[b_{1,17}]$. Suppose $M$  is a multipacking on $G_1$ such that $|M|=\MP(G_{1})$. So, $|M\cap N_3[b_{1,7}]|\leq 3$ and $|M\cap N_2[b_{1,17}]|\leq 2$. Therefore, $|M\cap (N_3[b_{1,7}]\cup N_2[b_{1,17}])|\leq 5$. So, $|M\cap V(G)|\leq 5$, that implies $|M|\leq 5$. Let $M_1=\{b_{1,1},b_{1,7},b_{1,13},b_{1,18},b_{1,21}\}$. Since $|N_r[v]\cap M|\leq r$ for each vertex $ v \in V(G_{1}) $ and for every integer $ r \geq 1 $, so $M_1$ is a multipacking of size $5$. Then  $ 5=|M_1|\leq |M|$. So, $|M|=5$. Therefore, $\MP(G_{1})=5$. 
\end{proof}

 So, now we have $\MP(G_1)=5$. Using this fact we prove that $\MP(G_2)=9$.

\begin{lemma}\label{mpG2=9}  $\MP(G_{2})=9$.
\end{lemma}
\begin{proof} As mentioned before,  $B_i= G_k[\{b_{i,j}:1\leq j \leq 21\}]$, $1\leq i \leq 2$. So, $B_1$ and $B_2$ are two blocks in $G_2$ which are isomorphic to $G_1$. Let $M$ be a multipacking of $G_2$ with size $\MP(G_{2})$. So, $|M|\geq 9$ by Lemma \ref{mpG2k>=9k}. Since $M$ is a multipacking of $G_2$, so $M\cap V(B_1)$  and $M\cap V(B_2)$  are multipackings of $B_1$ and $B_2$,  respectively. Let $M\cap V(B_1)=M_1$  and $M\cap V(B_2)=M_2$. Since $B_1\cong G_1$ and $B_2\cong G_1$, so $\MP(B_1)=5$ and $\MP(B_2)=5$ by Lemma \ref{mpG1=5}. This implies $|M_1|\leq 5$ and $|M_2|\leq 5$. Since $V(B_1)\cup V(B_2)=V(G_2)$ and $V(B_1)\cap V(B_2)=\phi$, so $M_1\cap M_2=\phi$ and $|M|=|M_1|+|M_2|$.  Therefore, $9\leq |M|=|M_1|+|M_2|\leq  10$. So, $9\leq |M|\leq 10$. 

We establish this lemma by using contradiction on $|M|$. In the first step, we prove that if  $|M_1|= 5$, then the particular vertex $b_{1,21}\in M_1$. Using this, we can show that $|M_2|\leq 4$. In this way we show that $|M|\leq 9$.

For the purpose of contradiction, we assume that $|M|=10$. So, $|M_1|+ |M_2|=10$, and also   $|M_1|\leq 5$, $|M_2|\leq 5$. Therefore,   $|M_1|=|M_2|= 5$.

\vspace{0.2cm}
\noindent
\textbf{Claim \ref{mpG2=9}.1. } If $|M_1|= 5$, then $b_{1,21}\in M_1$.
\begin{claimproof}
Suppose $b_{1,21}\notin M$. Let $S=\{b_{1,7},b_{1,14}\}$, $S_1=\{b_{1,r}:1\leq r\leq 6\}$, $S_2=\{b_{1,r}:8\leq r\leq 13\}$, $S_3=\{b_{1,r}:15\leq r\leq 20\}$. If $u,v\in S_t$, then $d(u,v)\leq 2$, this holds  for each $ t\in\{1,2,3\}$. So, by Lemma \ref{duv>=3}, $u,v$ together cannot be in a multipacking. Therefore $|S_t\cap M_1|\leq 1$ for $t=1,2,3$  and  $|S\cap M_1|\leq |S|= 2$. Now, $5=|M_1|=|M_1\cap [V(G_1)\setminus \{b_{1,21}\}|=|M_1\cap(S\cup S_1\cup S_2\cup S_3)|=|(M_1\cap S)\cup(M_1\cap S_1)\cup(M_1\cap S_2)\cup(M_1\cap S_3)|\leq |M_1\cap S|+|M_1\cap S_1|+|M_1\cap S_2|+|M_1\cap S_3|\leq 2+1+1+1=5$. Therefore,  $|S_t\cap M_1|= 1$ for $t=1,2,3$  and  $|S\cap M_1|= 2$, so $b_{1,7},b_{1,14}\in M_1$. Since $|S_2\cap M_1|= 1$,  there exists $ w \in S_2\cap M_1$. Then $N_2[b_{1,10}]$ contains three vertices $b_{1,7},b_{1,14},w$ of $M_1$, which is not possible. So, this is a contradiction. Therefore, $b_{1,21}\in M_1$.  
\end{claimproof}

\vspace{0.1cm}
\noindent
\textbf{Claim \ref{mpG2=9}.2. } If $|M_1|=5$, then $|M_2|\leq 4$.
\begin{claimproof}

Let $S'=\{b_{2,14},b_{2,21}\}$, $S_4=\{b_{2,r}:1\leq r\leq 6\}$, $S_5=\{b_{2,r}:8\leq r\leq 13\}$, $S_6=\{b_{2,r}:15\leq r\leq 20\}$. By  Lemma \ref{duv>=3},   $|S_t\cap M_2|\leq 1$ for $t=4,5,6$  and also $|S'\cap M_2|\leq |S'|= 2$.

Observe that, if $S_4\cap M_2\neq \phi$, then  $b_{2,7} \notin M_2$ (i.e.  if $b_{2,7} \in M_2$, then $S_4\cap M_2 = \phi$). [Suppose not, then $S_4\cap M_2 \neq \phi$ and  $b_{2,7} \in M_2$, so,  there exists $ u\in S_4\cap M_2$. Then $N_2[b_{2,3}]$ contains three vertices $b_{1,21},b_{2,7},u$ of $M$, which is not possible. This is a contradiction].

Suppose $S_4\cap M_2 \neq \phi$, then  $b_{2,7} \notin M_2$. Now, $5=|M_2|=|M_2\cap [V(B_2)\setminus \{b_{2,7}\}]|=|M_2\cap(S'\cup S_4\cup S_5\cup S_6)|=|(M_2\cap S')\cup(M_2\cap S_4)\cup(M_2\cap S_5)\cup(M_2\cap S_6)|\leq |M_2\cap S'|+|M_2\cap S_4|+|M_2\cap S_5|+|M_2\cap S_6|\leq 2+1+1+1=5$. Therefore  $|S_t\cap M_2|= 1$ for $t=4,5,6$  and  $|S'\cap M_2|= 2$. Since $|M_2\cap S_6|= 1$, so there exists $ u_1\in M_2\cap S_6$. Then $N_2[b_{2,17}]$ contains three vertices $b_{2,14},b_{2,21},u_1$ of $M_2$, which is not possible. So, this is a contradiction. 

Suppose $S_4\cap M_2 = \phi$, then either  $b_{2,7} \in M_2$ or $b_{2,7} \notin M_2$. First consider  $b_{2,7} \notin M_2$, then $5=|M_2|=|M_2\cap(S'\cup S_5\cup S_6)|=|(M_2\cap S')\cup(M_2\cap S_5)\cup(M_2\cap S_6)|\leq |M_2\cap S'|+|M_2\cap S_5|+|M_2\cap S_6|\leq 2+1+1=4$. So, this is a contradiction. And if $b_{2,7} \in M_2$, then  $5=|M_2|=|M_2\cap(S'\cup S_5\cup S_6\cup \{b_{2,7}\})|=|(M_2\cap S')\cup(M_2\cap S_5)\cup(M_2\cap S_6)\cup(M_2\cap \{b_{2,7}\})|\leq |M_2\cap S'|+|M_2\cap S_5|+|M_2\cap S_6|+|M_2\cap \{b_{2,7}\}|\leq 2+1+1+1=5$. Therefore  $|S_t\cap M_2|= 1$ for $t=5,6$  and  $|S'\cap M_2|= 2$. Since $|M_2\cap S_6|= 1$, so there exists $ u_2\in M_2\cap S_6$. Then $N_2[b_{2,17}]$ contains three vertices $b_{2,14},b_{2,21},u_2$ of $M_2$, which is not possible. So, this is a contradiction. So, $|M_1|=5\implies |M_2|\leq 4$.  
\end{claimproof}

Recall that for contradiction,  we assume $|M|=10$, which implies $|M_2|=5$. In the proof of the above claim, we established  $|M_2|\leq 4$, which in turn contradicts our assumption. So, $|M|\neq 10$.   Therefore, $|M|=9$. 
\end{proof}

Notice that  graph $G_{2k}$ has $k$ copies of $G_2$. Moreover, we have $\MP(G_2)=9$. Using the Pigeonhole principle, we show that $\MP(G_{2k})=9k$.

\begin{lemma}\label{mpG2k=9k} $\MP(G_{2k})=9k$, for each positive integer $k$.
\end{lemma}

\begin{proof} For $k=1$ it is true by Lemma \ref{mpG2=9}. Moreover, we know $\MP(G_{2k})\geq 9k$  by Lemma \ref{mpG2k>=9k}.  Suppose $k>1$ and assume  $\MP(G_{2k})>9k$. Let $\hat{M}$ be a multipacking of $G_{2k}$ such that $|\Hat{M}|>9k$. Let $\hat{B}_j$ be a subgraph of $G_{2k}$ defined as $\hat{B}_j=B_{2j-1}\cup B_{2j}$ where $1\leq j \leq k$. So, $V(G_{2k})=\bigcup_{j=1}^kV(\hat{B}_j)$ and $V(\hat{B}_p)\cap V(\hat{B}_q)=\phi$ for all $p\neq q$ and  $p,q\in \{1,2,3,\dots,k\}$. Since $|\Hat{M}|>9k$, so by the Pigeonhole principle there exists a  number  $j\in \{1,2,3,\dots,k\}$ such that $|\Hat{M}\cap \hat{B}_j|>9$. Since $\Hat{M}\cap \hat{B}_j$ is a multipacking of $\hat{B}_j$, so $\MP(\hat{B}_j)>9$.  But $\hat{B}_j \cong G_2$ and $\MP(G_2)=9$ by Lemma \ref{mpG2=9}, so $\MP(\hat{B}_j)=9$, which is a contradiction. Therefore, $\MP(G_{2k})=9k$.
\end{proof}

R. C. Brewster and  L. Duchesne \cite{brewster2013broadcast} introduced fractional multipacking in 2013 (also see \cite{teshima2014multipackings}).  Suppose $G$ is a graph with $V(G)=\{v_1,v_2,v_3,\dots,v_n\}$ and $w:V(G)\rightarrow [0,\infty)$  is a function. So, $w(v)$ is a weight on a vertex $v\in V(G)$. Let $w(S)=\sum_{u\in S}w(u)$ where $S\subseteq V(G)$.  We say   $w$ is a \textit{fractional multipacking} of $G$,  if $w( N_r[v])\leq r$ for each vertex $ v \in V(G) $ and for every integer $ r \geq 1 $. The \textit{fractional multipacking number} of $ G $ is the  value $\displaystyle \max_w w(V(G)) $ where $w$ is any fractional multipacking and it
	is denoted by $ \MP_f(G) $. A \textit{maximum fractional multipacking} is a fractional multipacking $w$  of a graph $ G $ such that	$ w(V(G))=\MP_f(G)$. If $w$ is a fractional multipacking, we define   a vector $y$ with the entries $y_j=w(v_j)$.  So,  $$\MP_f(G)=\max \{y.\mathbf{1} :  yA\leq c, y_{j}\geq 0\}.$$  So, this is a  linear program which is the dual of the linear program   $\min \{c.x :  Ax\geq \mathbf{1}, x_{i,k}\geq 0\}$. Let,  $$\gamma_{b,f}(G)=\min \{c.x : Ax\geq \mathbf{1}, x_{i,k}\geq 0\}.$$ Using the strong duality theorem for linear programming, we can say that  $$\MP(G)\leq \MP_f(G)= \gamma_{b,f}(G)\leq \gamma_{b}(G).$$

\begin{figure}[ht]
    \centering
   \includegraphics[height=11.3cm]{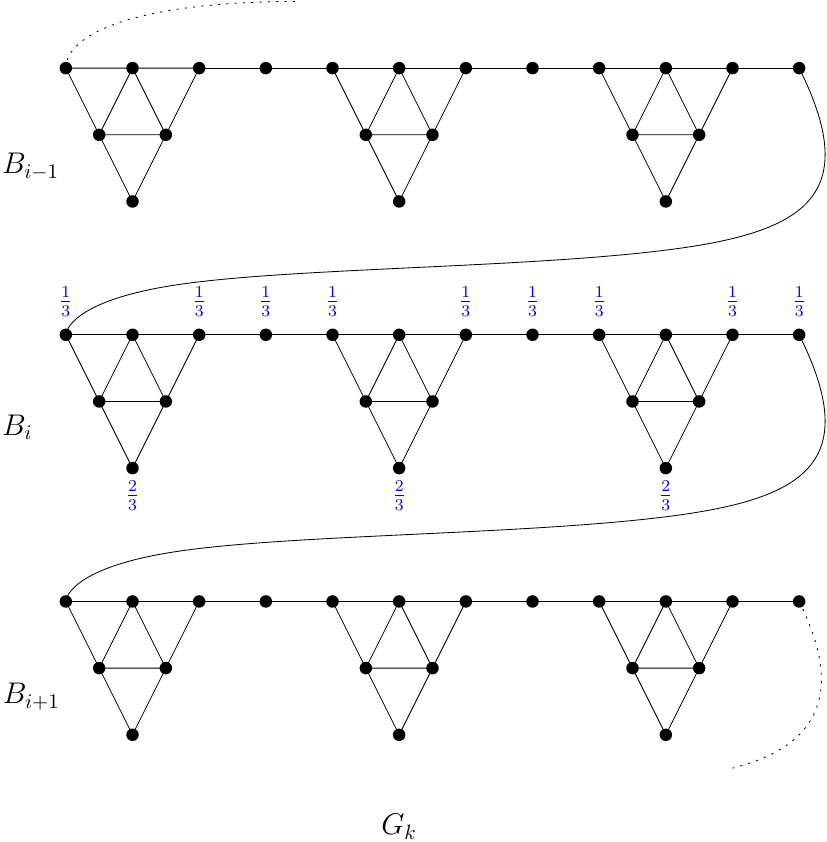}
    \caption{Graph $G_k$}
    \label{fig:Gkmpf}
\end{figure}

\begin{lemma} \label{mpfGk>=5k}
If $k$ is a positive integer, then $ \MP_f(G_{k})\geq 5k$.
\end{lemma}

\begin{proof}
We define a function $w: V(G_k)\rightarrow [0,\infty)$  where $w(b_{i,1})=w(b_{i,6})=w(b_{i,7})=w(b_{i,8})=w(b_{i,13})=w(b_{i,14})=w(b_{i,15})=w(b_{i,20})=w(b_{i,21})=\frac{1}{3}$ and $w(b_{i,4})=w(b_{i,11})=w(b_{i,18})=\frac{2}{3}$ for each $i\in \{1,2,3,\dots,k\}$ (Figure~\ref{fig:Gkmpf}).  So, $w(G_k)=5k$. 
 We want to show that $w$ is a fractional multipacking of $G_{k}$. So, we have to prove that $w(N_r[v])\leq r$ for each vertex $ v \in V(G_{k}) $ and for every integer $ r \geq 1 $. We prove this statement using induction on $r$.	It can be checked that $w(N_r[v])\leq r$ for each vertex $ v \in V(G_{k}) $ and for each  $ r \in \{1,2,3,4\} $. Now assume that the statement is true for $r=s$, we want to prove that it is true for $r=s+4$. Observe that, $w(N_{s+4}[v]\setminus N_{s}[v])\leq 4$, $\forall v\in V(G_{k})$. Therefore,  $w(N_{s+4}[v])\leq w(N_{s}[v])+4\leq s+4$. So, the statement is true. So, $w$ is a fractional multipacking of $G_{k}$. Therefore, $\MP_f(G_{k})\geq 5k$. 
\end{proof}

\begin{lemma} \label{mpfGk=gammabGk=5k}
If $k$ is a positive integer, then $\MP_f(G_{k})=\gamma_b(G_{k})= 5k$.
\end{lemma}

\begin{proof} Define a broadcast ${f}$ on $G_k$ as ${f}(b_{i,j})=
    \begin{cases}
        2 & \text{if } 1\leq i \leq k \text{ and } j=6,17\\
        1 & \text{if } 1\leq i \leq k \text{ and } j=12 \\
        0 & \text{otherwise }
    \end{cases}$.\\
Here  ${f}$ is an efficient dominating broadcast and $\sum_{v\in V(G_k)}{f}(v)=5k$. So, $\gamma_b(G_k)\leq 5k$, $\forall k\in \mathbb{N}$. So, by the strong duality theorem and Lemma \ref{mpfGk>=5k},  $5k\leq \MP_f(G_k)= \gamma_{b,f}(G_k)\leq \gamma_{b}(G_k)\leq 5k$. Therefore, $\MP_f(G_{k})=\gamma_b(G_{k})= 5k$. 
\end{proof}

So, $\gamma_b(G_{2k})=10k$ by Lemma \ref{mpfGk=gammabGk=5k} and  $\MP(G_{2k})=9k$ by Lemma \ref{mpG2k=9k}. Take $G_{2k}=H_k$. Thus we prove Theorem \ref{thm:9k10k}.


\begin{corollary} \label{mpfG-mpG}  The difference $\MP_f(G)-\MP(G)$ can be arbitrarily large for  connected  chordal graphs.
\end{corollary}

\begin{proof}
    We get $\MP_f(G_{2k})=10k$ by Lemma \ref{mpfGk=gammabGk=5k} and  $\MP(G_{2k})=9k$ by Lemma \ref{mpG2k=9k}. Therefore,   $\MP_f(G_{2k})-\MP(G_{2k})=k$  for all positive integers $k$. 
\end{proof}

\begin{corollary}  \label{gammabG2k/mpG2k}
For every integer $k \geq 1$, there is a connected chordal graph $G_{2k}$ with $\MP(G_{2k})=9k$, $\MP_f(G_{2k})/\MP(G_{2k})=10/9$ and $\gamma_b(G_{2k})/\MP(G_{2k})=10/9$.
\end{corollary}


\section{Hyperbolic graphs}
\label{sec:A study of Broadcast domination and Multipacking numbers on Hyperbolic graphs}
In this section, we relate the broadcast domination and multipacking number of $\delta$-hyperbolic graphs. In addition, we provide an approximation algorithm for the multipacking problem of the same. 

 Chepoi et al.~\cite{chepoi2008diameters} established a relation between the length of the radius and diameter of a $\delta$-hyperbolic graph, that we state in the following theorem:

\begin{theorem}[\cite{chepoi2008diameters}]\label{delta_hyperbolic_radius_diameter}
    For any $\delta$-hyperbolic graph $G$, we have $diam(G)\geq 2 \rad(G)-4\delta-1$. 
    
\end{theorem}

Using this theorem, we can establish a relation between $\gamma_{b}(G)$ and $\MP(G)$ of a $\delta$-hyperbolic graph $G$.

\deltaMultipackingBroadcastRelation*

\begin{proof} Let $G$ be a $\delta$-hyperbolic graph with radius $r$ and diameter $d$. 
    From Theorem \ref{d+1/3leqmpG}, $\big\lceil{\frac{d+1}{3}\big\rceil}\leq \MP(G)$  which implies that $d \leq 3\MP(G)-1$. Moreover, from  Theorem \ref{mpGleqgammabG} and Theorem \ref{delta_hyperbolic_radius_diameter},  $\gamma_{b}(G)\leq r \leq \big\lfloor{\frac{d+4\delta +1}{2}\big\rfloor} \leq \big\lfloor{\frac{(3\MP(G)-1)+4\delta +1}{2}\big\rfloor}$ $=\big\lfloor{\frac{3}{2} \MP(G)+2\delta \big\rfloor}  $. Therefore, $\gamma_{b}(G)\leq \big\lfloor{\frac{3}{2} \MP(G)+2\delta\big\rfloor}$. 
\end{proof}

\approxdeltaMultipacking*

\begin{proof}
    The proof is similar to the proof of Propostition \ref{prop:3/2mpGapprox}. If $P=v_0,\dots,v_d$ is a diametrical path of $G$, then the set $M=\{v_i:i\equiv 0 \text{ } (mod \text{ } 3), i=0,1,\dots,d\}$ is a multipacking of $G$ of size $\big\lceil{\frac{d+1}{3}\big\rceil}$ by Theorem \ref{d+1/3leqmpG}. We can construct $M$ in polynomial-time since we can find a diametral path of a graph $G$ in polynomial-time.   Theorem \ref{d+1/3leqmpG}, Theorem \ref{mpGleqgammabG} and Theorem \ref{delta_hyperbolic_radius_diameter} yield that  $\big\lceil{\frac{2\MP(G)-4\delta}{3}\big\rceil}\leq\big\lceil{\frac{2r-4\delta}{3}\big\rceil}\leq\big\lceil{\frac{d+1}{3}\big\rceil}\leq \MP(G)$. 
\end{proof}

\section{Conclusion}
\label{sec:Conclusion}

We have shown that the  bound $\gamma_b(G)\leq 2\MP(G)+3$ for general graphs $G$ can be improved to $\gamma_{b}(G)\leq \big\lceil{\frac{3}{2} \MP(G)\big\rceil}$ for connected  chordal graphs. It is known that for strongly chordal graphs, $\gamma_{b}(G)=\MP(G)$, we have shown that this is not the case for connected  chordal graphs. Even more, $\gamma_b(G)-\MP(G)$ can be arbitrarily large for connected chordal graphs, as we have constructed infinitely many connected chordal graphs $G$ where $\gamma_b(G)/ \MP(G)=10/9$ and $\MP(G)$ is arbitrarily large. Moreover, we have shown that $\gamma_{b}(G)\leq \big\lfloor{\frac{3}{2} \MP(G)+2\delta\big\rfloor} $ holds for all $\delta$-hyperbolic graphs. Note that connected split graphs have radius at most~2, so their broadcast number and their multipacking number are both at most~2. The graph from Figure~\ref{fig:S3} is a split graph, showing that the multipacking number of a split graph can indeed be different from its broadcast domination number.


We naturally ask for the best possible ratio on chordal graphs, which our work leaves open. This problem could also be studied for other interesting graph classes, such as the one in Figure~\ref{fig:diagram}, or others.

For general connected graphs, although the optimal ratio is now known asymptotically since $\displaystyle{\lim_{\MP(G)\to \infty}\sup\left\{\frac{\gamma_{b}(G)}{\MP(G)}\right\}=2}$, with a lower bound coming from hypercubes~\cite{Rajendraprasad25}, it remains an interesting open problem to tighten this ratio further. One direction is to show that $\gamma_b(G)\leq 2\MP(G)$ holds. For the lower bound, it would be nice to have constructions of connected graphs $G$ with arbitrarily large values of $\MP(G)$ that attain the ratio of~2. In fact, the exact values of $\MP(H_d)$ are still open~\cite{Rajendraprasad25} and, perhaps, they come close to a ratio of~2?

\section*{Declarations}

\noindent\textbf{Funding:} This research was financed by the IFCAM project ``Applications of graph homomorphisms'' (MA/IFCAM/18/39). Florent Foucaud was financed by the French government IDEX-ISITE initiative 16-IDEX-0001 (CAP 20-25), the International Research Center "Innovation Transportation and Production Systems" of the I-SITE CAP 20-25, the ANR project GRALMECO (ANR-21-CE48-0004), and the CNRS IRL RELAX during the preparation of this manuscript. 








\bibliographystyle{elsarticle-num}
\bibliography{ref}
\end{document}